\documentstyle[11pt]{article}

\begin{document}
\topmargin 0pt
\oddsidemargin 5mm
\setcounter{page}{1}
\begin{titlepage}
\hfill Preprint YeRPHI-1563(12)-2000

\vspace{2cm}
\begin{center}

{\bf  Long-Range Interactions of the Ball}\\
\vspace{5mm}
{\large R. A. Alanakyan}\\
\vspace{5mm}
{(C) All Rights Reserved \\}
\vspace{5mm}
{\em Theoretical Physics Department,
Yerevan Physics Institute,
Alikhanian Brothers St.2,
Yerevan 375036, Armenia\\}
 {E-mail: alanak@lx2.yerphi.am\\}

\vspace{5mm}
\vspace{5mm}
\vfill
\end{center}
\centerline{{\bf{Abstract}}}
In model independent way considered long-range interactions of the ball under
 assumption that any particle of this objects interact with probe particle as $ A/r^n$. 
Also presented  model-independent corrections to the Coulomb energy levels from regularised
 version of the potential $ A/r^n$.

\vspace{5mm}
\vfill
\begin{center}
{\bf Yerevan Physics Institute\\}
\end{center}

\end{titlepage}               
In this article considered long-range interactions of the ball under assumption that the 
unit of the its volume   interact with probe particle as $ A/r^n$.
There are many examples of the  such singular potentials: e.g. Van-der-Vaals 
interactions.  Also must be noticed that 
in some of the  theories with higher dimensions (see  e.g. \cite{GRS1},\cite{RS} and references
 therein)  appear the following corrections to Newton potential \cite{C1}:
\begin{equation}
\label{A1}
U(r)=\frac{Gm_1m_2}{r}(1+A/r^n)
\end{equation}
Previously has been considered the following corrections to Newton potential which also
 appear in some of multidimensional theories (see references in \cite{S}):
\begin{equation}
\label{A2}
U(r)=\frac{Gm_1m_2}{r}(1+se^{-Mr})
\end{equation}
  In number of papers  (see references in \cite{S}) has been calculated the potential of the
 ball with homogeneous density under assumption that
any particle of the ball interact with probe particle in accordance with formula (2). In \cite{S}
has been also published the experimental restriction on  parameters $s$ and$M$.
Analogously, using data of the  experiments described in  \cite{S} in principle possible to obtain some
 restriction on  $A$ and $n$.

We present also model-independent corrections to the energy levels  of the electron in the Coulomb field 
of the nuclei from regularized version of the potential   $ A/r^n$.

   We obtain for potential of ball with constant density :
\begin{equation}
\label{A3}
U_n(r)=\int d^3r_1\frac{A}{|\vec{r}-\vec{r'}|^n}=
-\frac{2\pi A}{r}\int r'dr'(\frac{|r-r'|^{(2-n)}}{2-n}- \frac{(r+r')^{(2-n)}}{2-n})
\end{equation}
At $r>R$  we have after integration:
\begin{equation}
\label{A4}
U_n(r)=\frac{2\pi A}{r}\frac{1}{n-2}
[\frac{(r-R)^{4-n}}{4-n}-\frac{r(r-R)^{3-n}}{3-n}-\frac{(r+R)^{4-n}}{4-n}
+\frac{r(r+R)^{3-n}}{3-n}]
\end{equation}
From (17) it is seen that potential is singular at   $r \rightarrow R$ ($U(r)\sim |r-R|^{(4-n)}$.
 
From this formula it is seen that cases $n=2,3,4$ must be considered separately.We obtain:
\begin{equation}
\label{A5}
U_2(r)=\frac{2\pi A}{r}(rR+\frac{1}{2}(R^2-r^2)log(\frac{r+R}{|r-R|}),
\end{equation}
\begin{equation}
\label{A6}
U_3(r)=\frac{2\pi A}{r}[r log(\frac{r+R}{r-R})-2R],   
\end{equation}
\begin{equation}
\label{A7}
U_4(r)=\frac{\pi A}{r}[log(\frac{(r-R)}{r+R})+\frac{2rR}{r^2-R^2}].   
\end{equation}
During derivation of $U_2(r)$ has been taken into account potential $U_n(r)$ at  $r<R$ (see formula (11) below). 
From (4) for $n=5,6,7$ we obtain the following simplifications:
\begin{equation}
\label{A8}
U_5(r)=\frac{4\pi AR^3}{3r}\frac{1}{(r^2-R^2)^2  } 
\end{equation}
\begin{equation}
\label{A9}
U_6(r)=\frac{4\pi AR^3}{3}\frac{1}{(r^2-R^2)^3}   
\end{equation}
\begin{equation}
\label{A10}
U_7(r)=\frac{4\pi AR^3}{15r}\frac{(5r^2+R^2)}{(r^2-R^2)^4}   
\end{equation}
At $n=1$  we obtain as it must be $U_1(r)=4\pi R^3 A/3r$ .
At $r<R$ and  $n<3$ we have:
\begin{equation}
\label{A11}
U_n(r)=\frac{2\pi A}{r}\frac{1}{n-2}
[\frac{(R-r)^{4-n}}{4-n}+\frac{r(R-r)^{3-n}}{3-n}-\frac{(r+R)^{4-n}}{4-n}
+\frac{r(r+R)^{3-n}}{3-n}]
\end{equation}

At  $n \geq 3$ integral $\int r'dr'\frac{1}{(r-r')^{(2-n)}}$ is divergent if   $r<R$.

It mean that inside ball ($r<R$) and $n \geq 3$ we must regularised
 interaction  $ A/r^n$.
We choose the following regularizations:
\begin{equation}
\label{A12}
V_n(r)=\frac{A}{(r^2+a^2)^{n/2}}
\end{equation}
\begin{equation}
\label{A13}
V_n(r)=\frac{A}{(r+a)^n},
\end{equation}
\begin{equation}
\label{A14}
V_n(r)=\frac{A}{r(r+r_0)^n}.
\end{equation}
where$r_0, a<<R$. For example $a$ may be a size of the molecules (atoms)
 which contained in this ball.The last potential transformed into Coulomb potential at  $r<<r_0$.

Using regularized potential (12) we obtain for $n=5$ the following result:

\begin{equation}
\label{A15}
V_5(r)=\frac{2\pi A}{5r}[-f_-+f_++\frac{r}{a^2}((R-r)f_-+(R+r)f_+)]   
\end{equation}
where  $f_{\pm}=((R\pm r)^2+a^2)^{-\frac{1}{2}}$
At  $r>R$ and  $|r-R| >> a$ we again obtain $A/r^{n}$, and at  $r<R$ ,$r, |r-R|>>a$ we have:
\begin{equation}
\label{A16}
V_5(r)=\frac{4\pi A}{5}[\frac{1}{a^2}-\frac{1}{R^2-r^2}].  
\end{equation}
Now we consider in model independent way (the nature of correction may be any including cases 
which has been considered above)
corrections to the energy levels  of the electron in the Coulomb field 
of the nuclei from regularised version of the potential  $ A/r^n$
expressed by formula (14):
\begin{equation}
\label{A22}
\delta E=\int d^3r |\psi| ^2 A/(r(r+r_0)^n)\approx 4 A a_B^{-3}  r_0^{2-n}\frac{2n-3}{(n-2)(n-1)}  
\end{equation}
Here has been taken into account that  at $r_0<<a_B$ in integral we can put $\psi^2=1/\pi a_B^3$ at $l=0$.
Analogously, in case of potential (13) we obtain:
\begin{equation}
\label{A23}
\delta E=\int d^3r |\psi| ^2 A/((r+r_0)^n)\approx 8A  a_B^{-3} r_0^{3-n}\frac{1}{(n-3)(n-2)(n-1)}.  
\end{equation}
This formulas are valid if $n>3$.
At $n=3$ by consideration of  two range in integral ($0<r<L$, $L<r< \infty$ where   $r_0<<L<<a_B$ ) 
e.g. from (22) we obtain:
\begin{equation}
\label{A22}
\delta E=\int d^3r |\psi| ^2 A/((r+r_0)^3)\approx 4\pi A(1/3-2 +log(a_B/2r_0)-C)
 \end{equation}
where $C$-is Eiler constant.

  The author express his sincere gratitude to G.V.Grigoryan for helpful
 discussions.

\end{document}